\documentclass[prd,aps,preprint,amsmath,amssymb,eshowkeys,nofootinbib]{revtex4-2}
\usepackage[mathscr]{euscript}
\usepackage{dcolumn}
\usepackage{multirow}
\usepackage{bm}
\usepackage{graphicx}
\usepackage{subfigure}
\usepackage{rotating}
\usepackage{amsmath,amssymb,amsthm}
\usepackage[colorlinks=true,linkcolor=red,urlcolor=green,citecolor=green]{hyperref}
\newcommand{\bea}{\begin{eqnarray}}
\newcommand{\eea}{\end{eqnarray}}
\newcommand{\beq}{\begin{equation}}
\newcommand{\eeq}{\end{equation}}

\def\/{\over}

\begin{document}

\title{Observational constraints on fractional holographic dark energy in the light of DESI DR2}
\author{Qihong Huang$^{1}$\footnote{Corresponding author: huangqihongzynu@163.com}, Yuchen Zhang$^{2}$, Bing Xu$^{3}$, and Kaituo Zhang$^{4}$}

\affiliation{
$^1$ School of Physics and Electronic Science, Zunyi Normal University, Zunyi, Guizhou 563006, China\\
$^2$ Department of Physics, Key Laboratory of Low Dimensional Quantum Structures and Quantum Control of Ministry of Education, and Hunan Research Center of the Basic Discipline for Quantum Effects and Quantum Technologies, Hunan Normal University, Changsha, Hunan 410081, China\\
$^3$ School of Electrical and Electronic Engineering, Anhui Science and Technology University, Bengbu, Anhui 233030, China\\
$^4$ Department of Physics, Anhui Normal University, Wuhu, Anhui 241000, China
}

\begin{abstract}
Based on the fractional entropy from fractional quantum mechanics, fractional holographic dark energy (FHDE) has been proposed with the Hubble horizon as the IR cutoff (FHDEH). We extend this framework by adopting the future event horizon and the particle horizon as the IR cutoff, proposing the FHDEF and FHDEP models. Using the SN+OHD+DESI DR2 dataset to constrain these models, we find that all three models provide a marginally lower $\chi^{2}_{min}$ compared to $\Lambda$CDM but without significant preference according to AIC and BIC. When CMB distance priors are included, the FHDEH and FHDEP models are strongly ruled out. We further analyze the cosmological evolution for these models, and find that only the FHDEF model predicts nearly identical evolutions of $\Omega_{m}$ and $\Omega_{de}$ to those of the $\Lambda$CDM model across cosmic history, but its deceleration parameter $q$ deviate from the $\Lambda$CDM model in the future, indicating richer late time dynamics beyond the standard $\Lambda$CDM cosmology. 

\textbf{Keywords}: holographic dark energy; observational constraints; DESI DR2
\end{abstract}

\maketitle

\section{Introduction}

Many cosmological observations suggest that the universe is experiencing an accelerated expansion~\cite{Perlmutter1999, Riess1998, Spergel2003, Spergel2007, Tegmark2004, Eisenstein2005}. A natural way to explain this remarkable phenomenon is to introduce dark energy~\cite{Peebles2003}, with the cosmological constant representing the simplest theoretical candidate and providing an excellent fit to a wide range of observational data~\cite{Planck2020}. However, the cosmological constant model faces some problems~\cite{Weinberg1989, Steinhardt1999}. To alleviate these problems, holographic dark energy (HDE) has been proposed and has attracted lots of attention~\cite{Cohen1999, Hsu2004, Horvat2004, Li2004, Wang2017}, its cosmological implications have been investigated from both observational constraints~\cite{Zhang2005,Li2009,Li2025d,Li2025e} and dynamical evolution~\cite{Zhang2005a,Huang2009}. Its framework is based on the holographic principle motivated by black hole thermodynamics and on the connection between the ultraviolet cutoff of a quantum field theory and the largest distance scale of the theory. Within the HDE framework, the choice of horizon entropy is crucial, since it uniquely determines the energy density, giving rise to various HDE models~\cite{Wang2017}. For instance, the original HDE model arises from taking the Bekenstein–Hawking entropy as the horizon entropy~\cite{Cohen1999, Hsu2004, Li2004}, while Tsallis HDE models are based on Tsallis entropy as a generalization of the standard Boltzmann–Gibbs entropy~\cite{Tavayef2018}, and Barrow HDE models originate from Barrow's modification of the Bekenstein–Hawking entropy formula incorporating quantum gravitational corrections with a fractal horizon structure~\cite{Saridakis2020}. Recently, fractional holographic dark energy (FHDE) has been proposed~\cite{Trivedi2024}, motivated by fractional entropy arising from fractional quantum mechanics in the context of Schwarzschild black hole thermodynamics~\cite{Jalalzadeh2021}, and it successfully describes the late time acceleration of the universe.

Beyond theoretical consistency, observational constraints are crucial for assessing the viability of HDE models, necessitating comparison with precise data, such as the Pantheon+ SN Ia sample~\cite{Riess2022, Brout2022, Scolnic2022} and observational Hubble parameter data (OHD)~\cite{Cao2022}. As standardizable candles, SN Ia allow for accurate distance measurements and were pivotal in revealing the late time acceleration of the universe. In parallel, OHD, inferred from galaxy ages or baryon acoustic oscillations, provide direct insights into the expansion history. As a result, researchers routinely combine SN Ia and OHD datasets as a powerful means of constraining cosmological parameters~\cite{Shen2025, Wang2025, Liu2024, Liu2024b, Arora2024, Oliveros2024, Wang2023, Mukherjee2022a, Mukherjee2022, Cao2021a, Pacif2021, Cao2021, Akarsu2020, Yang2020, Liu2019, Jimenez2016, Gong2013, Su2011, Gong2010, Wu2010a, Gong2006, Wu2006}. Recently, the Dark Energy Spectroscopic Instrument (DESI) collaboration released the baryon acoustic oscillation measurements from its second data release (DR2)~\cite{Abdul2025}, representing the largest spectroscopic galaxy sample to date. The DESI DR2 BAO measurements serve as a standard probe of the cosmic expansion history and are now widely used to constrain the dynamical properties of dark energy~\cite{Li2026a, Li2026b, Li2026c, Li2026d, Du2026, Ren2026, Wang2026a, Wang2026b, Du2025, Li2024d, Pang2025, Wang2025a, Yadav2026, Zhu2026, Plaza2025, Petri2026, Li2026, Li2025, Li2025a, Li2025b, Luciano2026, Wang2026}. These high precision datasets serve to both test the consistency of dark energy models and place stringent constraints on their key parameters. For the FHDE model, constraints from a combination of multiple observational datasets show that, with the model parameter $\alpha=1$, it not only achieves late time accelerated expansion but also yields an age of the universe consistent with that of the standard $\Lambda$CDM model~\cite{Robles-Barba2025}. However, in the original FHDE formulation, the fractional parameter $\alpha$ is theoretically constrained to $1 < \alpha \leq 2$. It remains an open question whether this theoretically favored range is consistent with observational data. In addition, different choices of the IR cutoff can lead to qualitatively different cosmological evolutions~\cite{Wang2017}. In this paper, we systematically constrain three FHDE models with different IR cutoffs, namely the Hubble horizon, the future event horizon, and the particle horizon, using current observational dataset and assess their observational viability.

The goals of this paper are twofold: first, to constrain three FHDE models with different IR cutoffs using observational data and identify which models are observationally viable; second, to analyze the cosmological evolution of the viable model, including the evolution of the cosmological parameters and phase space dynamics, to understand its late time behavior and attractor structure. This paper is structured as follows. Section~\ref{sec:2} introduces the FHDE models. Section~\ref{sec:3} presents the observational constraints on the model parameters using the Pantheon+ SN Ia+OHD+DESI DR2+CMB dataset. Section~\ref{sec:4} investigates the cosmological evolution of these FHDE models, as well as its phase space dynamics and attractor behavior. Our main conclusions are summarized in Section~\ref{sec:5}.

\section{Models} \label{sec:2}

Based on the holographic principle and the fractional entropy obtained from fractional quantum mechanics in the context of black hole thermodynamics~\cite{Jalalzadeh2021}, the FHDE energy density takes the form~\cite{Trivedi2024}
\begin{equation}
\rho_{de}=3 C^{2} L^{\frac{2-3\alpha}{\alpha}},\label{rhode0}
\end{equation}
where the fractional parameter $\alpha$ is constrained to $1 < \alpha \leq 2$ as derived from fractional quantum mechanics. This model reduces to the standard HDE for $\alpha=2$, to Barrow HDE for $\alpha=\frac{2}{\Delta+1}$, and to Tsallis HDE for $\alpha=\frac{2}{2\delta-1}$.

We consider a homogeneous and isotropic Friedmann--Robertson--Walker universe, whose line element is given by
\begin{equation}
ds^{2}=-dt^{2}+a^{2}(t)(dr^{2}+r^{2}d\Omega^{2}),
\end{equation}
and the Friedmann equation takes the form
\begin{equation}
H^{2}=\frac{\kappa^{2}}{3}\big(\rho_{r}+\rho_{m}+\rho_{de}\big),\label{H2}
\end{equation}
with $\rho_{r}$, $\rho_{m}$, and $\rho_{de}$ denoting the energy densities of radiation, pressureless matter, and FHDE, respectively, and obeying the conservation equations
\begin{eqnarray}
&& \dot{\rho}_{r}+4H\rho_{r}=0,\label{rhor}\\
&& \dot{\rho}_{m}+3H\rho_{m}=0,\label{rhom}\\
&& \dot{\rho}_{de}+3H(1+\omega_{de})\rho_{de}=0,\label{rhode}
\end{eqnarray}
where the equation of state parameter $\omega_{de}$ is defined as
\begin{equation}
\omega_{de}=\frac{p_{de}}{\rho_{de}}.
\end{equation}

After introducing the following dimensionless variables
\begin{equation}
\Omega_{r}=\frac{\kappa^{2}\rho_{r}}{3H^{2}}, \quad \Omega_{m}=\frac{\kappa^{2}\rho_{m}}{3H^{2}}, \quad \Omega_{de}=\frac{\kappa^{2}\rho_{de}}{3H^{2}},
\end{equation}
we can rewrite the Friedmann equation~(\ref{H2}) as
\begin{equation}
\Omega_{r}+\Omega_{m}+\Omega_{de}=1.\label{O1}
\end{equation}
Combining Eqs.~(\ref{H2})--(\ref{rhode}) and~(\ref{O1}) yields
\begin{equation}
\frac{\dot{H}}{H^{2}}=\frac{1}{2}\Big[\Omega_{m}+(1-3\omega_{de})\Omega_{de}\Big]-2.\label{HH2}
\end{equation}

Combining Eqs.~(\ref{rhom}),~(\ref{rhode}),~(\ref{O1}), and~(\ref{HH2}) with the definition $'=\frac{d}{d(\ln a)}$ yields the automatic dynamical equations governing this system
\begin{eqnarray}
&& \Omega'_{m}=[(3\omega_{de}-1)\Omega_{de}-\Omega_{m}+1]\Omega_{m},\label{Omm}\\
&& \Omega'_{de}=[(3\omega_{de}-1)(\Omega_{de}-1)-\Omega_{m}]\Omega_{de},\label{Omde}
\end{eqnarray}

The deceleration parameter $q$ is defined as
\begin{equation}
q=-1-\frac{\dot{H}}{H^{2}}.\label{q0}
\end{equation}

To constrain the model parameters using observational data and analyze the cosmological evolution based on the obtained constraints, we will analyze three FHDE models: (1) the original FHDE model with the Hubble horizon as the IR cutoff (FHDEH); (2) FHDE with future event horizon as IR cutoff (FHDEF); (3) FHDE with particle horizon as IR cutoff (FHDEP).

\subsection{FHDEH}\

With the Hubble horizon chosen as the IR cutoff, the FHDEH energy density ~(\ref{rhode0}) becomes~\cite{Trivedi2024}
\begin{equation}
\rho_{de}=3 C^{2} H^{\frac{3\alpha-2}{\alpha}}.\label{rhodeH}
\end{equation}
Using Eqs.~(\ref{rhode}), ~(\ref{HH2}), and ~(\ref{rhodeH}), the equation of state parameter $\omega_{de}$ can be solved as
\begin{equation}
\omega_{de}=\frac{(3\alpha-2)(\Omega_{m}+\Omega_{de})+8-6\alpha}{3[(3\alpha-2)\Omega_{de}-2\alpha]}.\label{omegadeH}
\end{equation}

\subsection{FHDEF}

When the future event horizon is taken as the IR cutoff, the FHDEF energy density has the form
\begin{equation}
\rho_{de}=3 C^{2} R_{F}^{\frac{2-3\alpha}{\alpha}},\label{rhodeF}
\end{equation}
with
\begin{equation}
R_{F}=a\int^{\infty}_{t}\frac{dt}{a},
\end{equation}
which satisfies the relation $\dot{R}_{F}=H R_{F}-1$. By combining Eqs.~(\ref{rhode}), ~(\ref{HH2}), and ~(\ref{rhodeF}), the equation of state parameter $\omega_{de}$ can be derived as
\begin{equation}
\omega_{de}=-\frac{1}{3\alpha}\Big[(3\alpha-2)F+2\Big],\label{omegadeF}
\end{equation}
with
\begin{equation}
F=\frac{1}{H R_{F}}=\frac{1}{\mathcal{C}^{\frac{\alpha}{2-3\alpha}}}, \quad \mathcal{C}=\frac{1}{\kappa^{2} C^{2}} H^{\frac{2-\alpha}{\alpha}}\Omega_{de}
\end{equation}
which satisfies
\begin{equation}
F'=-\Big[ \frac{1}{2}(\Omega_{m}+(1-3\omega_{de})\Omega_{de})-1-F \Big]F.\label{FF1}
\end{equation}

\subsection{FHDEP}

For the particle horizon as the IR cutoff, the FHDEP energy density takes the form
\begin{equation}
\rho_{de}=3 C^{2} R_{P}^{\frac{2-3\alpha}{\alpha}},\label{rhodeP}
\end{equation}
with
\begin{equation}
R_{P}=a\int^{t}_{0}\frac{dt}{a},
\end{equation}
which satisfies the relation $\dot{R}_{P}=H R_{P}+1$. With Eqs.~(\ref{rhode}), ~(\ref{HH2}), and ~(\ref{rhodeP}), the equation of state parameter $\omega_{de}$ can be expressed as
\begin{equation}
\omega_{de}=\frac{1}{3\alpha}\Big[(3\alpha-2)P-2\Big],\label{omegadeP}
\end{equation}
with
\begin{equation}
P=\frac{1}{H R_{P}}=\frac{1}{\mathcal{C}^{\frac{\alpha}{2-3\alpha}}}, \quad \mathcal{C}=\frac{1}{\kappa^{2} C^{2}} H^{\frac{2-\alpha}{\alpha}}\Omega_{de},
\end{equation}
which satisfies
\begin{equation}
P'=-\Big[ \frac{1}{2}(\Omega_{m}+(1-3\omega_{de})\Omega_{de})-1+P \Big]P.\label{PP1}
\end{equation}

\section{Observational Constraints} \label{sec:3}

In previous section, we introduces three different FHDE models. In this section, we constrain the model parameters using four observational datasets: Pantheon+ SN Ia sample covering the redshift range $z \in [0.001, 2.261]$~\cite{Riess2022, Brout2022, Scolnic2022}, OHD sample spanning $z \in [0.07, 1.965]$ collected by~\cite{Cao2022} as shown in Table~\ref{Tab0}, BAO measurements from DESI DR2 covering $z \in [0.295,2.33]$~\cite{Abdul2025}, and CMB distance priors from Planck 2018~\cite{Zhai2019,Chen2019}. 

\begin{table}[h]
\centering
\caption{\label{Tab0} OHD data~\cite{Cao2022}.}
\resizebox{\textwidth}{!}{
 \begin{tabular}{|c|c|c|c|c|c|c|c|}
  \hline
  \hline
  $z$ & $H(z)$[km s$^{-1}$ Mpc$^{-1}$] & $z$ & $H(z)$[km s$^{-1}$ Mpc$^{-1}$] & $z$ & $H(z)$[km s$^{-1}$ Mpc$^{-1}$] & $z$ & $H(z)$[km s$^{-1}$ Mpc$^{-1}$] \\
  \hline
  0.07 & 69.0$\pm$19.6 & 0.28 & 88.8$\pm$36.6 & 0.4783 & 80.9$\pm$9.0 & 0.9 & 117.0$\pm$23.0\\
  \hline
  0.09 & 69.0$\pm$12.0 & 0.352 & 83.0$\pm$14.0 & 0.48 & 97.0$\pm$62.0 & 1.037 & 154.0$\pm$20.0\\
  \hline
  0.12 & 68.6$\pm$26.2 & 0.3802 & 83.0$\pm$13.5 & 0.593 & 104.0$\pm$13.0 & 1.3 & 168.0$\pm$17.0\\
  \hline
  0.17 & 83.0$\pm$8.0 & 0.4 & 95.0$\pm$17.0 & 0.68 & 92.0$\pm$8.0 & 1.363 & 160.0$\pm$33.6\\
  \hline
  0.179 & 75.0$\pm$4.0 & 0.4004 & 77.0$\pm$10.2 & 0.75 & 98.8$\pm$33.6 & 1.43 & 177.0$\pm$18.0\\
  \hline
  0.199 & 75.0$\pm$5.0 & 0.4247 & 87.1$\pm$11.2 & 0.781 & 105.0$\pm$12.0 & 1.53 & 140.0$\pm$14.0\\
  \hline
  0.2 & 72.9$\pm$29.6 & 0.4497 & 92.8$\pm$12.9 & 0.875 & 125.0$\pm$17.0 & 1.75 & 202.0$\pm$40.0\\
  \hline
  0.27 & 77.0$\pm$14.0 & 0.47 & 89.0$\pm$50.0 & 0.88 & 90.0$\pm$40.0 & 1.965 & 186.5$\pm$50.4\\
  \hline
  \hline
  \end{tabular}
  }
\end{table}

\subsection{Data and Methodology}

For these FHDE models, we can write the expansion rate function as
\begin{equation}
E(z)=\frac{H(z)}{H_{0}}=\sqrt{\Omega_{r,0}(1+z)^{4}+\Omega_{m,0}(1+z)^{3}+\Omega_{de,0}e^{3\int_{0}^{z}\frac{1+\omega_{de}}{1+z}dz}},
\end{equation}
where $\Omega_{r,0}+\Omega_{m,0}+\Omega_{de,0}=1$. The theoretical apparent magnitude $m_{th}$, which corresponds to the predicted observable of SN Ia, is formulated as
\begin{equation}
m_{th}=5 \log_{10} \Big( \frac{D_{L}(z)}{Mpc} \Big) + 25 + M,
\end{equation}
where $M$ is the absolute magnitude of SN Ia and $D_{L}(z)$ represents the luminosity distance, which is defined as
\begin{equation}
D_{L}(z)=c(1+z)\int^{z}_{0}\frac{dz}{H(z)}.
\end{equation}

To estimate the model parameters for these FHDE modes, we perform Markov Chain Monte Carlo (MCMC) sampling with the emcee library in Python. For the FHDEH model, we constrain six parameters $\{ H_{0}, \Omega_{m,0}, \Omega_{de,0}, \alpha, M, r_{d}\}$; for the FHDEF and FHDEP models, we constrain seven parameters $\{ H_{0}, \Omega_{m,0}, \Omega_{de,0}, \alpha, \mathcal{C}, M, r_{d}\}$. The total log-likelihood is defined as
\begin{equation}
\ln(\mathcal{L}_{total})=-\frac{1}{2}\chi^{2}_{total}+const.,
\end{equation}
with
\begin{equation}
\chi^{2}_{total}=\chi^{2}_{SN}+\chi^{2}_{OHD}+\chi^{2}_{DESI}+\chi^{2}_{CMB}.
\end{equation}
Here, the $\chi^{2}$ term for the SN Ia sample is given by
\begin{equation}
\chi^{2}_{SN}=\left( \hat{m}_{obs} - m_{th} \right)^\dag C_{SN}^{-1} \left( \hat{m}_{obs} - m_{th} \right),
\end{equation}
where $\hat{m}_{obs}$ denotes the array of observed corrected apparent magnitude, and $C_{SN}$ represents the associated covariance matrix. 

For the OHD sample, the $\chi^{2}$ term is evaluated as
\begin{equation}
\chi^{2}_{OHD} = \sum_{i=1}^{N_{OHD}} \left(\frac{H_{obs,i}-H_{th}(z_i)}{\sigma_{OHD,i}}\right)^2
\end{equation}
where $H_{obs,i}$ and $\sigma_{OHD,i}$ denote the $i$-th observed value and its standard deviation, respectively, and $N_{OHD}$ is the total number of OHD data. 

For the DESI DR2 sample, the $\chi^{2}$ term is defined as
\begin{equation}
\chi^{2}_{DESI} = \sum_{i=1}^{N_{DESI}} \Delta_{i}^{T} \mathbf{C}_{i}^{-1} \Delta_{i},
\end{equation}
where the form of $\Delta_{i}$ and $\mathbf{C}_{i}$ depends on the measurement type provided for the $i$-th redshift bin. For anisotropic measurements, they are given by
\begin{eqnarray}
\Delta_{i}=
\left(
\begin{matrix}
D_{M}^{obs}(z_{i})/r_{d}-D_{M}^{th}(z_{i})/r_{d}\\[6pt]
D_{H}^{obs}(z_{i})/r_{d}-D_{H}^{th}(z_{i})/r_{d}
\end{matrix}
\right)
\end{eqnarray}
and
\begin{eqnarray}
\mathbf{C}_{i}=
\left(
\begin{matrix}
\sigma^{2}_{D_M/r_d} & r_{HM} \sigma_{D_M/r_d} \sigma_{D_H/r_d}\\
r_{HM} \sigma_{D_M/r_d} \sigma_{D_H/r_d} & \sigma^{2}_{D_H/r_d}
\end{matrix}
\right).
\end{eqnarray}
For isotropic measurements, $\Delta_{i}$ reduces to the scalar $D_{V}^{obs}(z_{i})/r_{d}-D_{V}^{th}(z_{i})/r_{d}$ with the corresponding variance $\sigma^{2}_{D_V/r_d}$. Here, the Hubble distance and the transverse comoving distance are defined as
\begin{equation}
D_{H}(z)=\frac{c}{H(z)},
\end{equation}
and
\begin{equation}
D_{M}(z)=c\int^{z}_{0}\frac{dz}{H(z)},
\end{equation}
and the volume averaged distance $D_{V}(z)$ used in the isotropic case is given by
\begin{equation}
D_{V}(z)=\left[ z D_{M}^{2}(z) D_{H}(z) \right]^{1/3}.
\end{equation}
Here, $\sigma_{D_{M}/r_{d}}$ and $\sigma_{D_{H}/r_{d}}$ represent the observational uncertainties associated to $D_{M}(z)/r_{d}$ and $D_{H}(z)/r_{d}$, respectively, $r_{HM}$ is their correlation coefficient between the two observables, and $r_{d}$ is the sound horizon during the drag epoch.

For the CMB distance priors, the $\chi^{2}$ term is given by
\begin{equation}
\chi^{2}_{CMB} = \Delta p^{T} \mathbf{C}_{CMB}^{-1} \Delta p, \qquad \Delta p=p^{obs}-p^{th},
\end{equation}
where $p=\{R,l_{A},\Omega_{b}h^{2}\}$, with $\mathbf{C}_{CMB}$ being the covariance matrix, and the CMB shift parameter $R$ and the acoustic scale $l_{A}$ take the form
\begin{equation}
R=\frac{D_{M}(z_{*})\sqrt{\Omega_{m}H^{2}_{0}}}{c}, \qquad l_{A}=\frac{\pi D_{M}(z_{*})}{r_{s}(z_{*})},
\end{equation}
where $z_{*}$ is the redshift at the photon decoupling epoch. According to Planck CMB observations, $p^{\text{obs}}=\{1.7502, 301.471, 0.02236\}$. Since $\Omega_{b}h^{2}$ correlates with $R$ and $l_{A}$, their covariance matrix, adopted from~\cite{Chen2019}, is incorporated into the $\chi^{2}$ calculation.

To compare the FHDE models with the standard $\Lambda$CDM model, we also perform a corresponding MCMC analysis of $\Lambda$CDM using the same observational datasets. Given the different numbers of parameters, we adopt the Akaike Information Criterion (AIC)~\cite{Akaike1974} and Bayesian Information Criterion (BIC)~\cite{Schwarz1978, Trotta2008} to statistically compare these four models, where
\begin{equation}
AIC=\chi^{2}_{min}+2n,
\end{equation}
and
\begin{equation}
BIC=\chi^{2}_{min}+n\ln(N),
\end{equation}
where $n$ is the number of parameters and $N$ is the number of observational data points. 

\subsection{Results and Discussion}

To estimate the model parameters, we perform MCMC sampling using the emcee library in Python 3.13.14. The theoretical predictions are computed using our independently developed Python code. The burn-in is determined by the integrated autocorrelation time, discarding twice the maximum autocorrelation time, and convergence is ensured by requiring the chain length to exceed 50 times the autocorrelation time for all parameters. The prior ranges adopted for these parameters are $H_{0}\in[50,80]$, $\Omega_{m}\in[0.1,0.5]$, $\Omega_{de}\in[0.5,0.9]$, $\alpha\in[1.0,2.0]$, $\mathcal{C}\in[0.0,2.0]$, $M\in[-20,-18]$, and $r_{d}\in[130,160]$.

\begin{figure}[h]
\begin{center}
\includegraphics[width=0.45\textwidth]{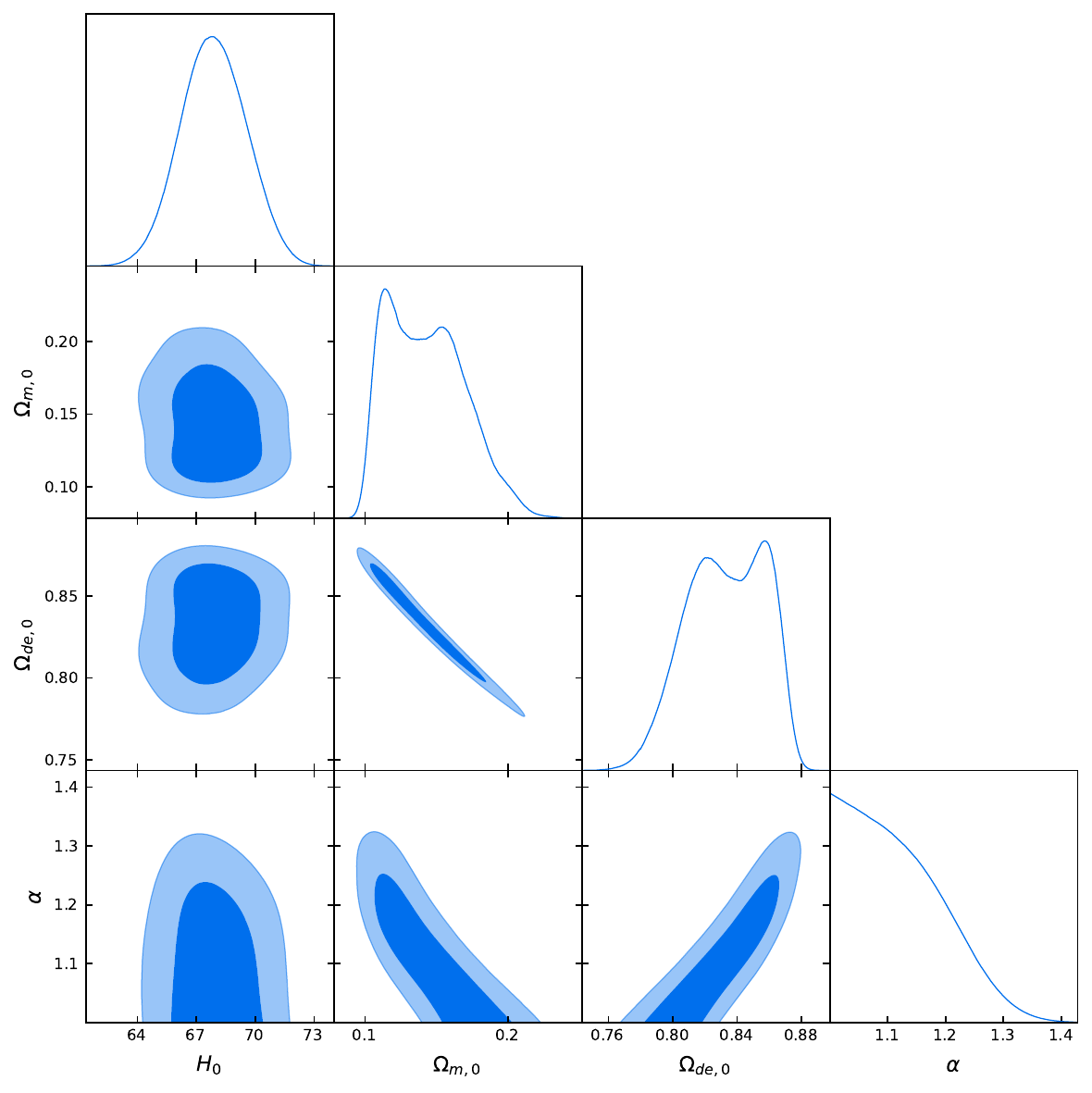}
\includegraphics[width=0.45\textwidth]{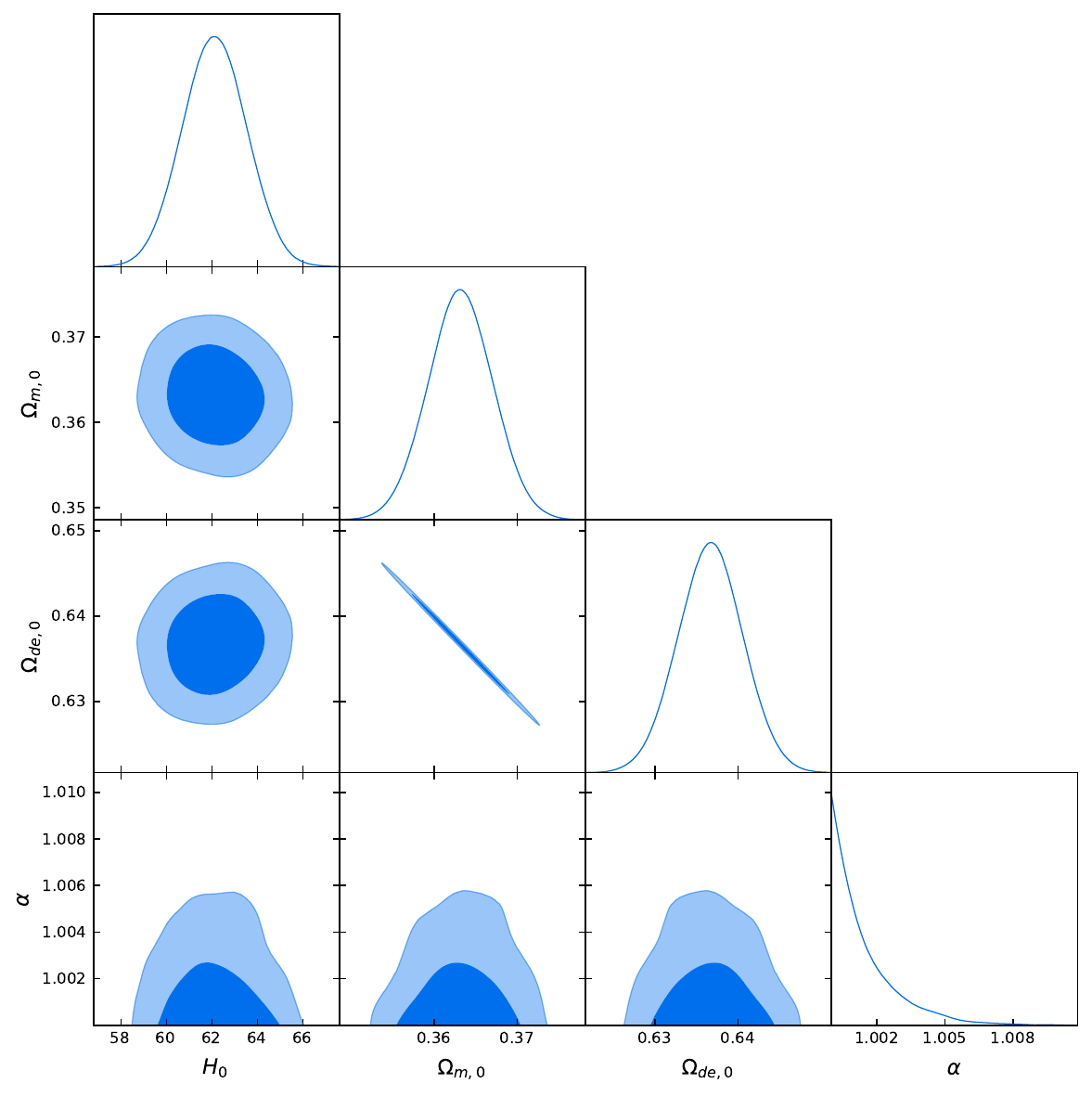}
\caption{\label{Fig1} Confidence contours for the parameters of the FHDEH model. The left panel uses SN, OHD, and DESI DR2 datasets, while the right panel uses SN, OHD, DESI DR2, and CMB distance priors. The units of $H_{0}$ are km s$^{-1}$ Mpc$^{-1}$.}
\end{center}
\end{figure}

\begin{figure}[h]
\begin{center}
\includegraphics[width=0.45\textwidth]{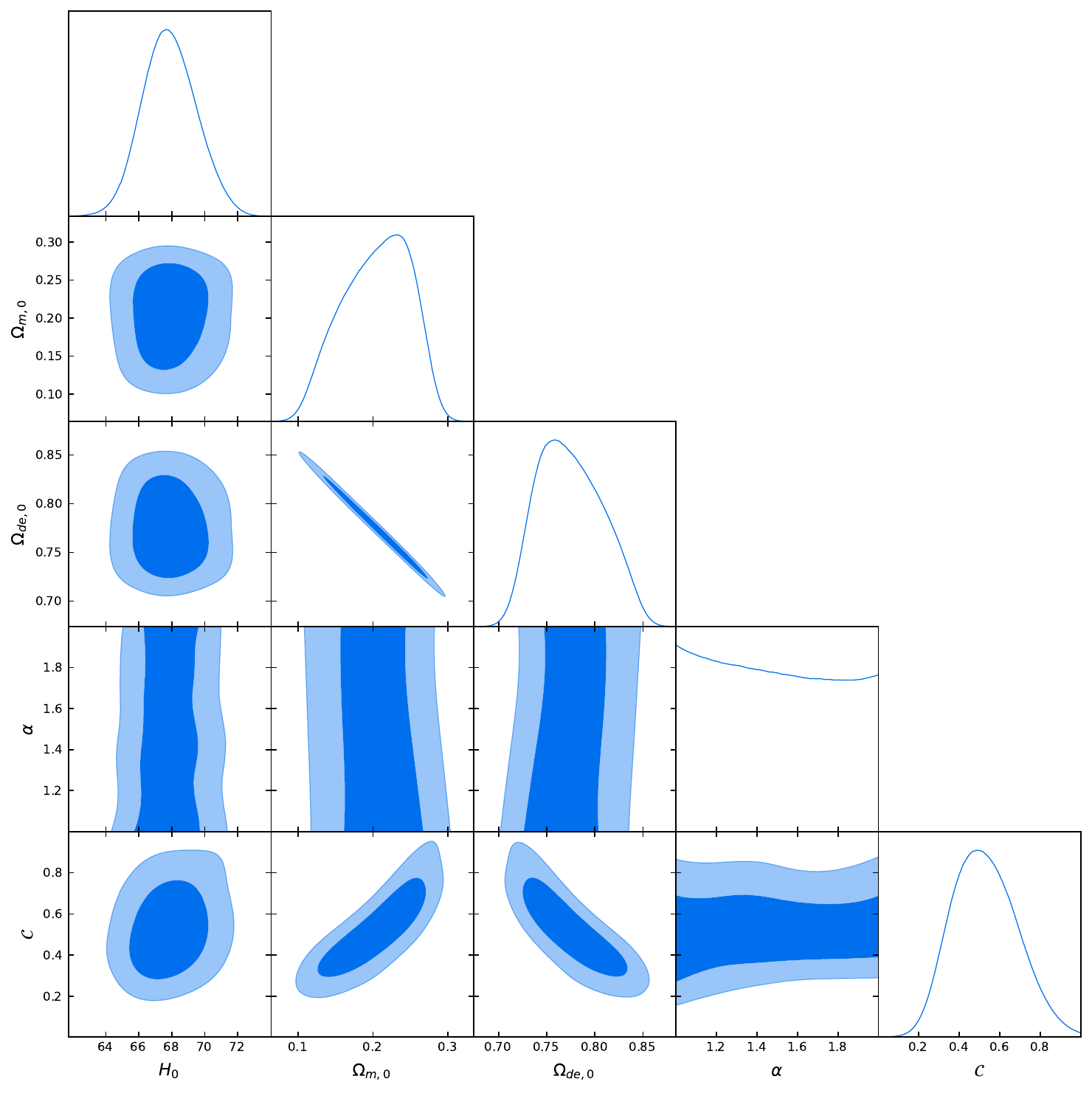}
\includegraphics[width=0.45\textwidth]{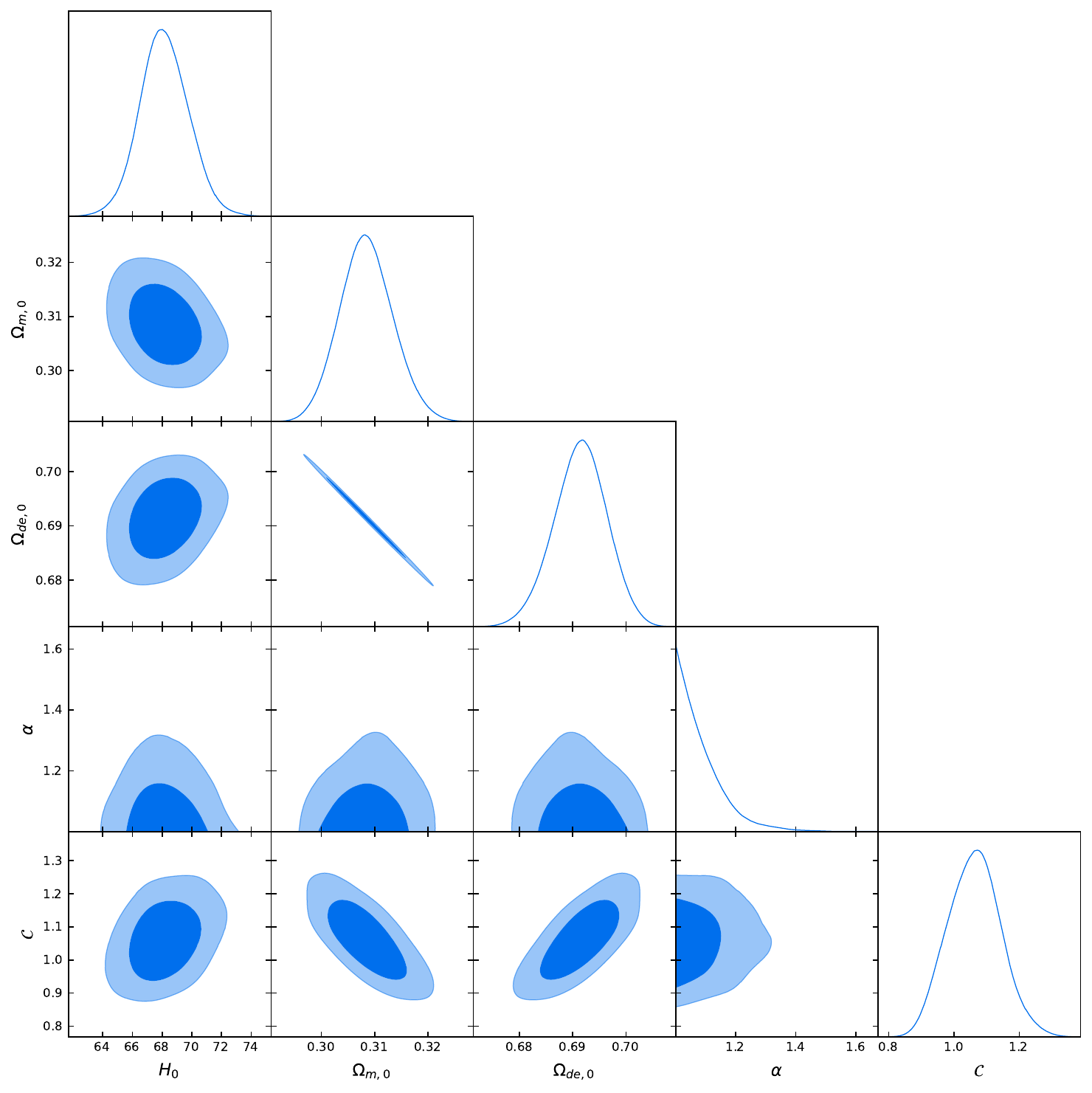}
\caption{\label{Fig2} Confidence contours for the parameters of the FHDEF model. The left panel uses SN, OHD, and DESI DR2 datasets, while the right panel uses SN, OHD, DESI DR2, and CMB distance priors. The units of $H_{0}$ are km s$^{-1}$ Mpc$^{-1}$.}
\end{center}
\end{figure}

\begin{figure}[h]
\begin{center}
\includegraphics[width=0.45\textwidth]{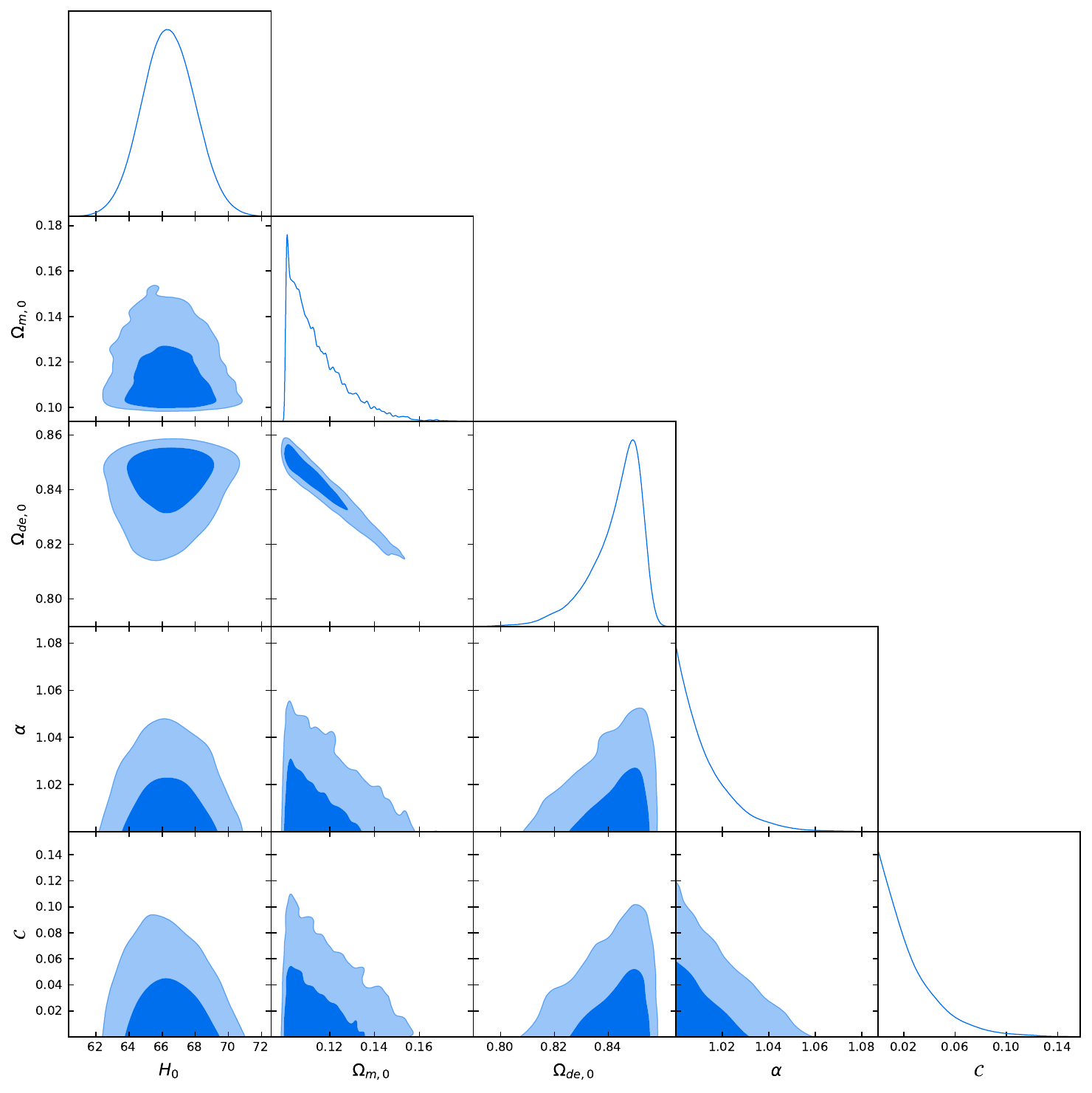}
\includegraphics[width=0.45\textwidth]{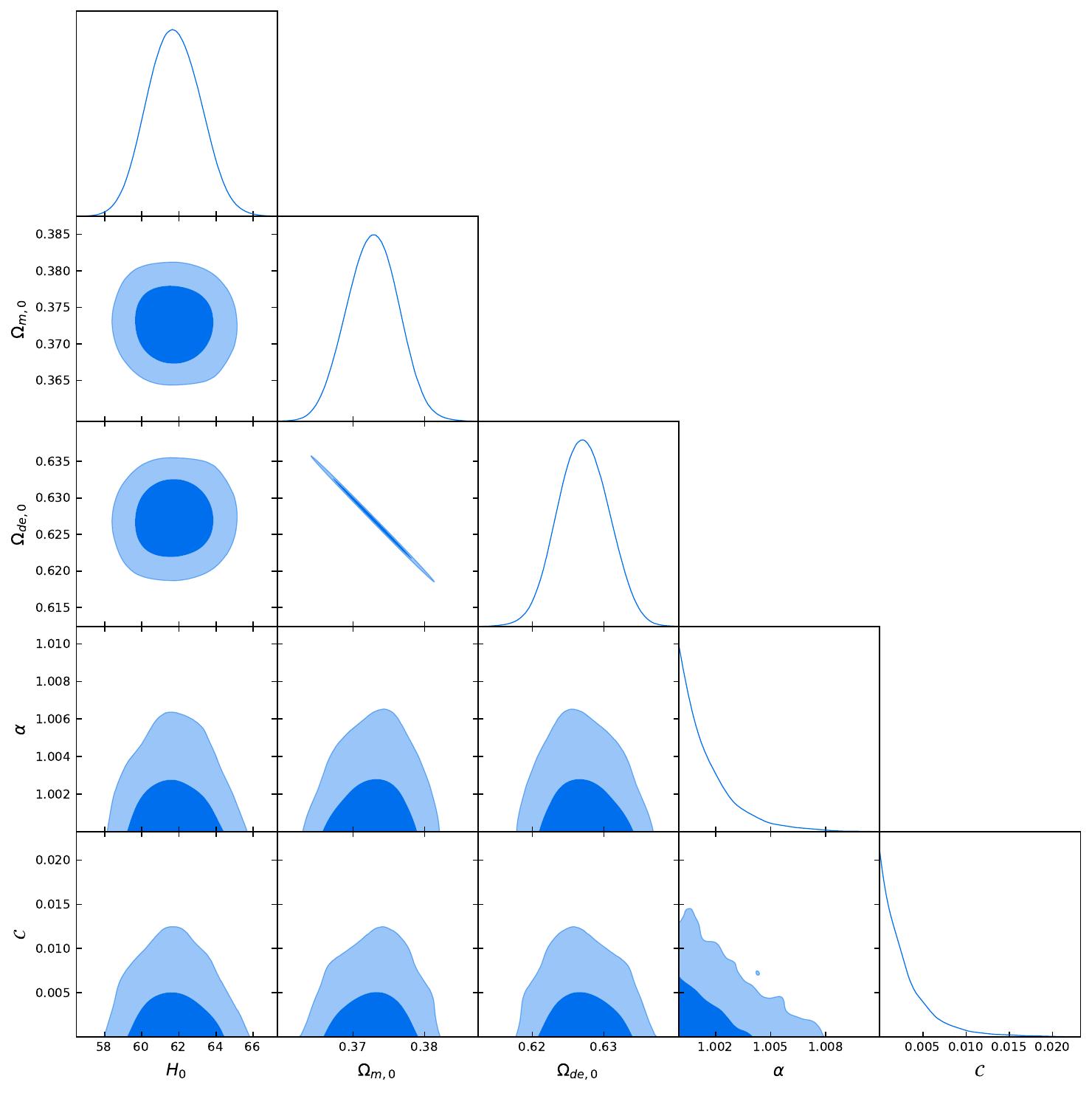}
\caption{\label{Fig3} Confidence contours for the parameters of the FHDEP model. The left panel uses SN, OHD, and DESI DR2 datasets, while the right panel uses SN, OHD, DESI DR2, and CMB distance priors. The units of $H_{0}$ are km s$^{-1}$ Mpc$^{-1}$.}
\end{center}
\end{figure}

We summarize the parameter constraints for the $\Lambda$CDM, FHDEH, FHDEF, and FHDEP models in Table~\ref{Tab1}, where the mean values and 1$\sigma$ confidence levels (CL) are listed. Three centered dots ($\cdots$) in this table indicate that the mean values of $\alpha$ cannot be calculated because the observational constraints are nearly flat across the entire range $1 < \alpha \leq 2$ derived from fractional quantum mechanics, indicating that all values in this interval are equally supported by the dataset. The posterior distributions for the FHDEH, FHDEF, and FHDEP models are presented in Figs.~\ref{Fig1}, ~\ref{Fig2}, and ~\ref{Fig3}, respectively. In these tables and figures, we adopt two datasets: SN+OHD+DESI DR2 and SN+OHD+DESI DR2+CMB. For the first dataset, the sound horizon at the drag epoch $r_{d}$ is treated as a free parameter. For the second dataset, $r_{d}$ is derived from the background model by the standard integration formula with the Planck 2018 best-fit value $z_{drag}=1059.94$, and the sound horizon at photon decoupling epoch $r_{*}$ is computed using the standard integration formula with the Planck 2018 best-fit value $z_{*}=1089.92$~\cite{Planck2020}.

\begin{table}[h]
\centering
\caption{\label{Tab1} Summary of observational constraints ($68\%$ CL) for $\Lambda$CDM and FHDE models from different data combinations. The quantities $\Delta\chi^{2}_{min}$, $\Delta AIC$, and $\Delta BIC$ are defined with respect to $\Lambda$CDM, thus negative values favor the FHDE models, while positive values favor $\Lambda$CDM. The units of $H_{0}$ and $r_{d}$ are km s$^{-1}$ Mpc$^{-1}$ and Mpc, respectively.}
\resizebox{\textwidth}{!}{
\begin{tabular}{|c|c|c|c|c|c|c|c|c|c|c|c|}
\hline
\hline
Datasets & Model & $H_{0}$ & $\Omega_{m,0}$ & $\Omega_{de,0}$ & $\alpha$ & $\mathcal{C}$ & $r_{d}$ & $\chi^{2}_{min}$ & $\Delta \chi^{2}_{min}$ & $\Delta AIC$ & $\Delta BIC$\\
\hline
\multirow{4}{*}{\shortstack{SN\\+OHD\\+DESI DR2}} & $\Lambda$CDM & $68.8^{+1.6}_{-1.7}$ & $0.296^{+0.012}_{-0.009}$ & $0.701^{+0.010}_{-0.009}$ & $-$ & $-$ & $147.1 \pm 3.3$ & $1431.5$ & $0$ & $0$ & $0$\\
\cline{2-12}
& FHDEH & $67.9 \pm 1.7$ & $0.142^{+0.019}_{-0.035}$ & $0.834^{+0.031}_{-0.022}$ & $<1.164$ & $-$ & $147.0^{+3.4}_{-3.5}$ & $1426.7$ & $-4.8$ & $-2.8$ & $2.6$\\
\cline{2-12}
& FHDEF & $67.9^{+1.5}_{-1.7}$ & $0.205^{+0.055}_{-0.039}$ & $0.775^{+0.029}_{-0.041}$ & $\cdots$ & $0.525^{+0.145}_{-0.176}$ & $147.0^{+3.4}_{-3.3}$ & $1426.8$ & $-4.7$ & $-0.7$ & $10.1$\\
\cline{2-12}
& FHDEP & $66.4 \pm 1.6$ & $0.114^{+0.004}_{-0.014}$ & $0.843^{+0.011}_{-0.005}$ & $<1.015$ & $<0.029$ & $147.1^{+3.5}_{-3.4}$ & $1430.9$ & $-0.6$ & $3.4$ & $14.2$\\
\hline
\multirow{4}{*}{\shortstack{SN\\+OHD\\+DESI DR2\\+CMB}} & $\Lambda$CDM & $68.8^{+1.5}_{-1.6}$ & $0.304 \pm 0.004$ & $0.695^{+0.003}_{-0.004}$ & $-$ & $-$ & $-$ & $1437.3$ & $0$ & $0$ & $0$\\
\cline{2-12}
& FHDEH & $62.1 \pm 1.4$ & $0.363 \pm 0.004$ & $0.637 \pm 0.004$ & $<1.002$ & $-$ & $-$ & $1637.5$ & $200.2$ & $202.2$ & $207.6$\\
\cline{2-12}
& FHDEF & $68.2^{+1.6}_{-1.7}$ & $0.309 \pm 0.005$ & $0.691 \pm 0.005$ & $<1.103$ & $1.062^{+0.077}_{-0.085}$ & $-$ & $1436.5$ & $-0.8$ & $3.2$ & $14$\\
\cline{2-12}
& FHDEP & $61.7^{+1.5}_{-1.4}$ & $0.373^{+0.003}_{-0.004}$ & $0.627 \pm 0.004$ & $<1.002$ & $<0.003$ & $-$ & $1687.8$ & $250.5$ & $254.5$ & $265.3$\\
\hline
\hline
\end{tabular}
}
\end{table}

For the $\Lambda$CDM model, both datasets constrain the mean value of $H_{0}$ to $68.8$km s$^{-1}$ Mpc$^{-1}$, and the minimum $\chi^{2}_{min}$ values are $1431.5$ and $1437.3$, respectively, indicating that the CMB distance priors are highly consistent with the low-redshift data within the $\Lambda$CDM framework, and adding the CMB distance priors does not shift the mean value.

When the dataset SN+OHD+DESI DR2 is used to constrain the FHDE models, for the FHDEH model, we obtain $H_{0}=67.9 \pm 1.7$km s$^{-1}$ Mpc$^{-1}$, which is slightly lower than that in the $\Lambda$CDM model, and the parameter $\alpha$ is constrained to be $\alpha<1.164$; for the FHDEF model, we find $H_{0}$ has the same mean value as the FHDEH model, the parameter $\alpha$ is unconstrained, and the parameter $\mathcal{C}$ is constrained as $\mathcal{C}=0.525^{+0.146}_{-0.176}$; for the FHDEP model, it yields the lowest $H_{0}$ among all models, and yields stringent upper limits $\alpha<1.015$ and $C<0.029$. The minimum $\chi^{2}_{min}$ values for FHDEH, FHDEF, and FHDEP are $1426.7$, $1426.8$, and $1430.9$, respectively, all lower than that of the $\Lambda$CDM model, indicating that the FHDE models provide a marginally lower $\chi^{2}_{min}$ than $\Lambda$CDM for this data combination. The $\Delta AIC$ values for FHDEH, FHDEF, and FHDEP are $-2.8$, $-0.7$, and $3.4$, respectively, the AIC of FHDEH and FHDEF are slightly lower than that of the $\Lambda$CDM model, indicating a marginal preference according to the AIC criterion. The $\Delta BIC$ values for FHDEH, FHDEF, and FHDEP are $2.6$, $10.1$, and $14.2$, respectively, all higher than that of the $\Lambda$CDM model, reflecting a penalty for the extra parameters $\alpha$ and $\mathcal{C}$. These results indicate that the BIC strongly favors the $\Lambda$CDM model over FHDEF and FHDEP models, and only slightly disfavors the FHDEH model.

When the dataset SN+OHD+DESI DR2+CMB, which includes the CMB distance priors, is used to constrain the FHDE models, for the FHDEH and FHDEP models, the parameter $\alpha$ is constrained to $\alpha<1.002$ and the values of $\Delta \chi^{2}_{min}$ exceed $200$, indicating that both models are strongly disfavored by the data compared to the $\Lambda$CDM model at a very high significance level; for the FHDEF model, we obtain $H_{0}=68.2^{+1.6}_{-1.7}$km s$^{-1}$ Mpc$^{-1}$, which is slightly lower than that in the $\Lambda$CDM model, and the parameters $\alpha$ and $\mathcal{C}$ are constrained to be $\alpha<1.103$ and $\mathcal{C}=1.062^{+0.077}_{-0.085}$. The values of $\Delta \chi^{2}_{min}$, $\Delta AIC$, and $\Delta BIC$ for FHDEF are $-0.8$, $3.2$, and $14$, respectively, suggesting that the FHDEF model provides a slightly lower $\chi^{2}_{min}$ than $\Lambda$CDM model, but the BIC strongly disfavors it compared to the $\Lambda$CDM model due to the penalty for extra parameters.

\begin{figure}[h]
\begin{center}
\includegraphics[width=0.6\textwidth]{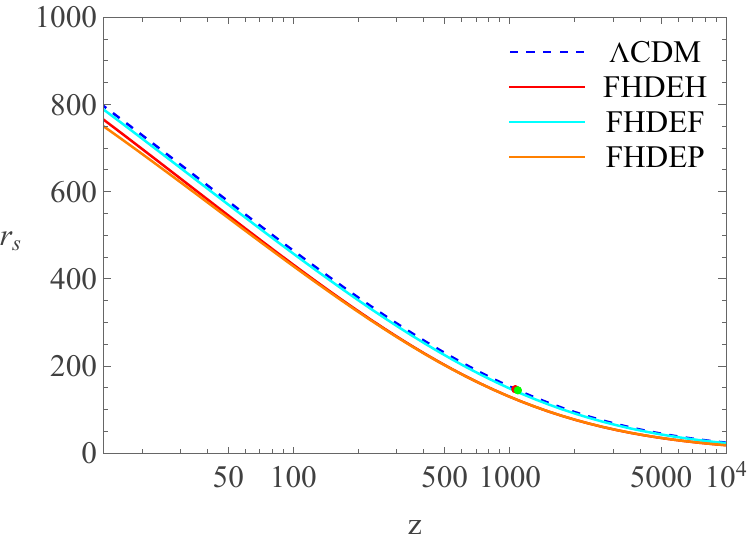}
\caption{\label{Fig4} Evolution of the comoving sound horizon $r_{s}$ as a function of redshift $z$. The red and green dots represent the Planck 2018 values at baryon drag and recombination epochs, respectively. The unit of $r_{s}$ is Mpc.}
\end{center}
\end{figure}

When comparing the constrained results from the datasets SN+OHD+DESI DR2 and SN+OHD+DESI DR2+CMB for the FHDEH and FHDEP models, we find when the CMB distance priors are introduced, $H_{0}$ is constrained to $H_{0}=62.1 \pm 1.4$km s$^{-1}$ Mpc$^{-1}$ and $H_{0}=61.7^{+1.5}_{-1.4}$km s$^{-1}$ Mpc$^{-1}$, both of which are substantially smaller than that of the $\Lambda$CDM model, reflecting that this worsens the Hubble tension compared to the $\Lambda$CDM model; the parameter $\alpha$ is constrained to $\alpha<1.002$ for both models, and the parameter $\mathcal{C}$ is constrained to $\mathcal{C}<0.003$ for the FHDEP model, both of which are very tight constraints; the values of $\Delta \chi^{2}_{min}$ for the FHDEH and FHDEP models are $200.2$ and $250.5$, demonstrating that both models severely deviate from the $\Lambda$CDM model. As shown in Fig.~\ref{Fig4}, the comoving sound horizon $r_{s}(z)$ of the FHDEH and FHDEP models deviates significantly from the $\Lambda$CDM model, with the deviation becoming more pronounced toward lower redshifts. These results demonstrate that the FHDEH and FHDEP models show acceptable fits to low-redshift data (SN+OHD+DESI DR2) alone, but they catastrophically fail when CMB distance priors (high redshift) are added, as evidenced by $\Delta \chi^{2}_{min}>200$ and the clear discrepancy in $r_{s}$ at $z \sim 1000$. This reveals that these models cannot reconcile the expansion history at recombination with that at late times, a capability that the $\Lambda$CDM model possesses. They are therefore strongly disfavored by the combined dataset, despite their low-redshift performance. However, in our previous work, when an interaction term is introduced to the FHDEH model, the resulting interacting FHDEH models yield a lower $\chi^{2}_{min}$ than that of the $\Lambda$CDM model~\cite{Huang2026}. Therefore, introducing an interaction term between the pressureless matter and the dark energy is worth considering as a possible resolution for the FHDEH and FHDEP models.

We note that the poor performance of the FHDEH and FHDEP models is primarily driven by the CMB distance priors rather than the SN data, as shown in Table~\ref{Tab1}. Therefore, although we have used the Pantheon+ sample, we do not expect that the qualitative conclusions would change with other SN Ia samples such as DESY5 or Union3, since the CMB constraints dominate.

In summary, using the SN+OHD+DESI DR2 dataset, FHDEH and FHDEF models provide fit statistics comparable to $\Lambda$CDM model, with FHDEH exhibiting the best fit among the three FHDE models. However, when the number of free parameters is taken into account, the $\Lambda$CDM model remains statistically favored overall. When the CMB distance priors are included in the SN+OHD+DESI DR2+CMB dataset, the FHDEH and FHDEP models are strongly disfavored, while the FHDEF model remains the only FHDE model that survives the inclusion of CMB distance priors, although its $\Delta BIC$ indicates that it is still substantially penalized compared to $\Lambda$CDM model.

\section{Evolution of the Universe} \label{sec:4}

In the previous section, we demonstrate that among the three FHDE models, only the FHDEF model survives the inclusion of CMB distance priors, while FHDEH and FHDEP models are strongly ruled out by the observational data. In this section, we focus on the evolutionary behavior of the universe for these FHDE models using the mean values of the parameters obtained from the dataset SN+OHD+DESI DR2+CMB. 

\subsection{Evolution of Cosmological Parameters}

\begin{figure*}[htp]
\begin{center}
\includegraphics[width=0.45\textwidth]{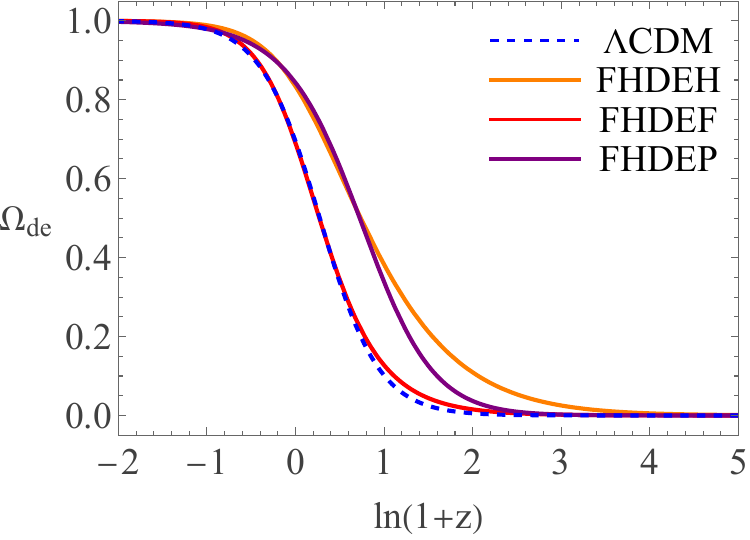}
\includegraphics[width=0.45\textwidth]{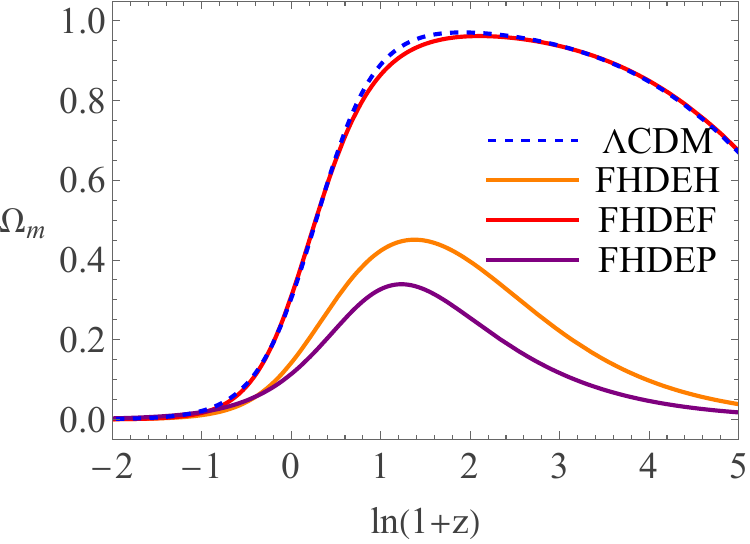}
\includegraphics[width=0.46\textwidth]{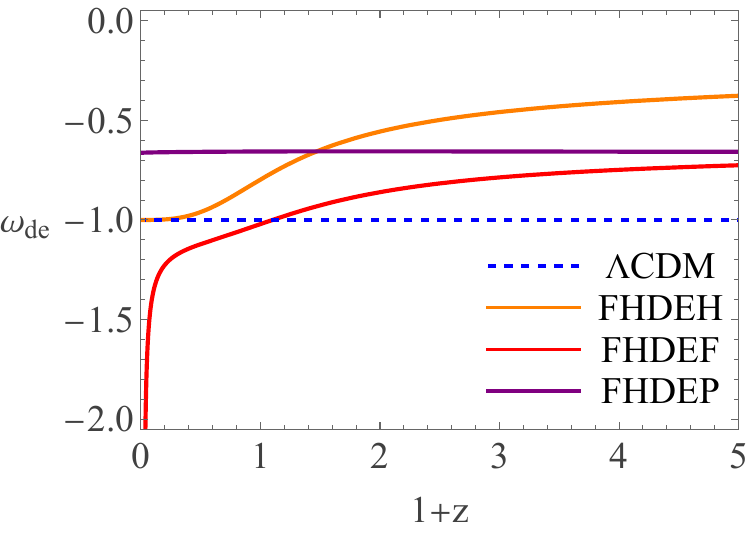}
\includegraphics[width=0.44\textwidth]{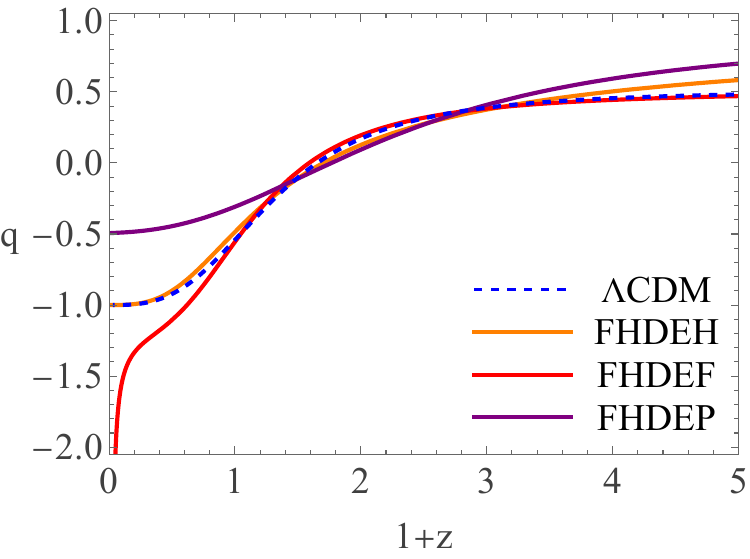}
\caption{\label{Fig5} Evolutionary curves of cosmological parameters for \textbf{the FHDE models using the mean values from the SN+OHD+DESI DR2+CMB dataset.}}
\end{center}
\end{figure*}

To analyze the evolution of cosmological parameters, we adopt the mean values from Table~\ref{Tab1} to solve the dynamical equation ~(\ref{Omm}), ~(\ref{Omde}), ~(\ref{FF1}), and~(\ref{PP1}) for these FHDE models. We then plot the evolutionary curves of $\Omega_{de}$, $\Omega_{m}$, $\omega_{de}$, and $q$ for these FHDE models, as shown in Fig.~\ref{Fig5}, in which we show the evolutionary curve for the $\Lambda$CDM model with blue dashed line and adopt the mean values for the $\Lambda$CDM model from Table~\ref{Tab1}.

In the first panel of Fig.~\ref{Fig5}, we can see that for all models, $\Omega_{de}$ approaches $0$ at early times and tends to $1$ at late times, indicating that FHDE models dominate the late time evolution of the universe. The second panel shows that $\Omega_{m}$ dominates at intermediate redshifts only for the FHDEF model; for the FHDEH and FHDEP models, $\Omega_{m}$ remains below $0.45$ and does not play a dominant role; in all cases, $\Omega_{m}$ approaches $0$ at late times. Overall, these two panels demonstrate that the FHDEF model exhibits only slight differences from the $\Lambda$CDM model in describing the evolution of $\Omega_{de}$ and $\Omega_{m}$. In contrast, the FHDEH and FHDEP models fail to reproduce a pressureless matter dominated epoch. The third panel shows that $\omega_{de}$ for the FHDEH and FHDEP models behaves as quintessence, but only the FHDEH model allow $\omega_{de}$ to approach $-1$ in the future; for the FHDEF model, $\omega_{de}$ behaves as quintessence in the past, has recently crossed the phantom divide, and currently lies in the phantom regime, indicating a quintom-like behavior that emerged at very late times. The fourth panel shows that the curves for the FHDEF and $\Lambda$CDM model overlap for $z>0$, indicating that they share similar deceleration behavior in the past and at the present epoch; for the FHDEH model, the curve overlaps with the $\Lambda$CDM model for $z<3$, but an obvious departure occurs for $z>3$; for the FHDEP model, the curve departs from the $\Lambda$CDM model significantly; for all models, a transition from deceleration to acceleration can be realized in the expected redshift region ($0.48 \leq z_{t} < 1$). These results indicate that the FHDEF model is dynamically closest to $\Lambda$CDM among the three FHDE models, while the FHDEH and FHDEP models exhibit significant deviations that are consistent with their poor performance in the observational constraints.

For the FHDEF model, when $1+z$ approaches to $0$, both $\omega_{de}$ and $q$ decrease rapidly from $-1.5$ to more negative values, as shown in Fig.~\ref{Fig5}, indicating that the phantom behavior becomes increasingly severe in the future. Such evolution inevitably leads to a divergence of the dark energy density $\rho_{de}$ and the Hubble parameter $H(z)$ within a finite cosmic time, which is known as the big rip singularity. This behavior is qualitatively different from that of the $\Lambda$CDM model; its future evolution points toward a dramatic fate, offering a clear distinction from the standard cosmological paradigm.

\begin{figure*}[htp]
\begin{center}
\includegraphics[width=0.6\textwidth]{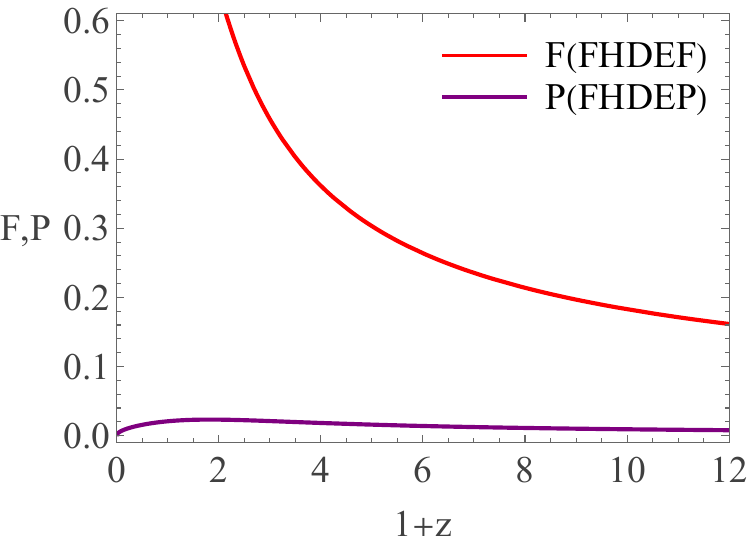}
\caption{\label{Fig51} Evolutionary curves of F for the FHDEF model and P for the FHDEP model using the mean values from the SN+OHD+DESI DR2+CMB dataset.}
\end{center}
\end{figure*}

In Fig.~\ref{Fig51}, adopting the mean values from the SN+OHD+DESI DR2+CMB dataset in Table~\ref{Tab1}, we plot the evolutionary curves of F for the FHDEF model and P for the FHDEP model. This figure shows that as $1+z$ approaches $0$, corresponding to the infinite future, $F$ increases, while $P$ decreases to $0$. This result means that, for the FHDEP model, $P$ approaches $0$ as $1+z$ approaches $0$, indicating that the particle horizon diverges in the infinite future.

In summary, the FHDEF model predicts evolutions of $\Omega_{de}$ and $\Omega_{m}$ that are nearly identical to those of $\Lambda$CDM across the entire cosmic history. The deceleration parameter $q$ also overlaps with that of $\Lambda$CDM for $z>0$, indicating that the two models are cosmologically indistinguishable in the past and at the present epoch. However, $\omega_{de}$ for FHDEF has recently crossed the phantom divide and currently lies in the phantom regime, which will drive a future evolution of $q$ that deviates from $\Lambda$CDM.

\subsection{Attractor Analysis}

In the previous subsection, we have analyzed the evolution of cosmological parameters, and found that the FHDEF model closely mimics $\Lambda$CDM in the past and present, while the FHDEH and FHDEP models exhibit significant deviations; the equation of state parameter $\omega_{de}$ for FHDEF model has recently crossed the phantom divide and entered the phantom regime, leading to a different future evolution. To further investigate this difference, we perform a dynamical system analysis of the attractor behavior for the FHDEF model. 

To achieve this goal, following Refs.~\cite{Bahamonde2018, Huang2019, Huang2021, Wu2010, Dutta2017, Wu2007, Wu2008, Huang2025a, Huang2025b}, we obtain the critical points by solving the autonomous system
\beq
\Omega_{m}'=\Omega_{de}'=F'=0.
\eeq
After solving Eqs.~(\ref{Omm}), ~(\ref{Omde}), and ~(\ref{FF1}), we obtain four critical points listed in Table~\ref{Tab2}. By linearizing the autonomous system, we derive a set of first order differential equations. The stability of the critical points is subsequently determined by the eigenvalues of the associated coefficient matrix. According to linear stability theory, a critical point is classified as an attractor if all eigenvalues are negative, as unstable if all are positive, and as a saddle point if the eigenvalues have mixed signs. Applying this criterion to the FHDEF model, we obtain the stability properties of the critical points which are summarized in Table~\ref{Tab2}.

\begin{table}[h]
\centering
\caption{\label{Tab2} Critical points and their stability for the FHDEF model.}
 \begin{tabular}{|c|c|c|c|c|c|c|c|}
  \hline
  \hline
  Label & $(\Omega_{m},\Omega_{de},F)$ & $\Omega_r$ & $\omega_{de}$ & $q$ & Eigenvalues & Conditions & Points \\
  \hline
  $P_{1}$ & $(0,0,0)$ & $1$ & $-\frac{2}{3\alpha}$ & $1$ & $(1,1,\frac{2+\alpha}{\alpha})$ & $1<\alpha\leq 2$ & Unstable\\
  \hline
  $P_{2}$ & $(1,0,0)$ & $0$ & $-\frac{2}{3\alpha}$ & $\frac{1}{2}$ & $(\frac{2}{\alpha},-1,\frac{1}{2})$ & $1<\alpha\leq 2$ & Saddle\\
  \hline
  $P_{3}$ & $(0,1,0)$ & $0$ & $-\frac{2}{3\alpha}$ & $\frac{1}{2}-\frac{1}{\alpha}$ & $(-\frac{2}{\alpha},\frac{1}{2}-\frac{1}{\alpha},-\frac{2+\alpha}{\alpha})$ & $1<\alpha< 2$ & Stable\\
  \hline
  $P_{4}$ & $(0,1,1)$ & $0$ & $-1$ & $-1$ & $(-\frac{1}{2}+\frac{1}{\alpha},-4,-3)$ & $1<\alpha< 2$ & Saddle\\
  \hline
  \hline
  \end{tabular}
\end{table}

As shown by the summarized results in Table~\ref{Tab2}, point $P_{1}$ denotes a radiation dominated decelerated epoch, $P_{2}$ represents a pressureless matter dominated decelerated epoch, $P_{3}$ and $P_{4}$ correspond to the dark energy dominated acceleration epoch, as indicated by their corresponding values of $q$; among these, only the value of $q$ for $P_{4}$ is equal to $-1$. For all points, the equation of state parameters $\omega_{de}$ are negative, and only the value of $\omega_{de}$ for $P_{4}$ reaches $-1$. Among these points, $P_{4}$ corresponds to the cosmological constant $\Lambda$, as it satisfies $q=-1$ and $\omega_{de}=-1$. The eigenvalue analysis further reveals that $P_{1}$ is an unstable node, $P_{2}$ and $P_{4}$ are saddle points, while $P_{3}$ is a stable attractor. This indicates that the FHDEF model admits a stable late time attractor $P_{3}$, consistent with the observed accelerated expansion, whereas the cosmological constant like point $P_{4}$ is only a saddle point and does not serve as the final evolutionary state of the universe.

\begin{figure}[h]
\begin{center}
\includegraphics[width=0.6\textwidth]{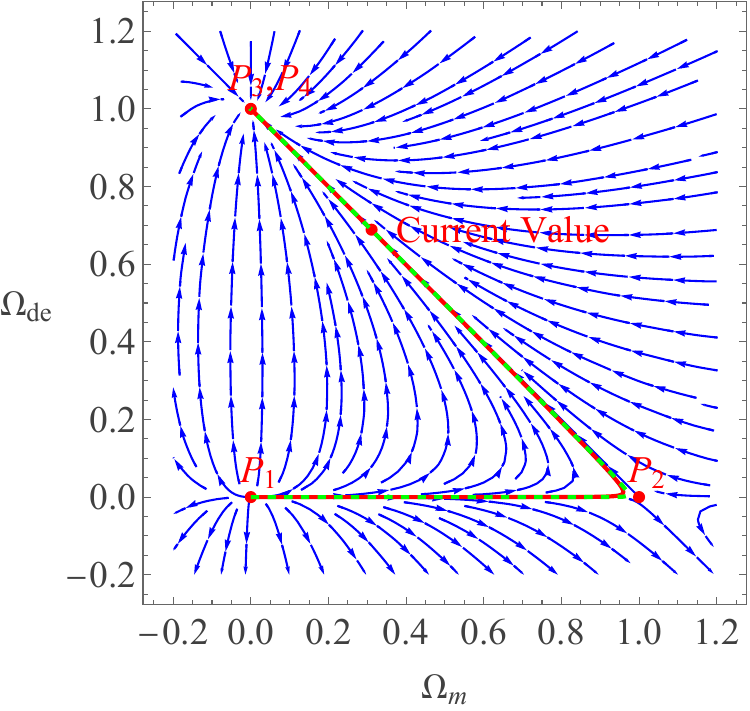}
\caption{\label{Fig6} Phase space trajectories in the $(\Omega_{m},\Omega_{de})$ plane for the $\Lambda$CDM and FHDEF models, shown as green dashed and red solid lines respectively, with mean parameter values taken from Table~\ref{Tab1}.}
\end{center}
\end{figure}

To analyze the evolution of the universe in the FHDEF model, we plot its phase space trajectories in the $(\Omega_{m},\Omega_{de})$ plane using the mean parameter values from Table~\ref{Tab1}, as shown in Fig.~\ref{Fig6}. Figure~\ref{Fig6} shows that the evolutionary trajectory for the FHDEF model overlaps with that of the $\Lambda$CDM model, and both evolve toward the same critical point $P_{3}/P_{4}(0,1)$. It should be noted that this point is a stable attractor for the $\Lambda$CDM model, while for the FHDEF model, $P_{3}$ is a stable point and $P_{4}$ is a saddle point.

\begin{figure}[h]
\begin{center}
\includegraphics[width=0.6\textwidth]{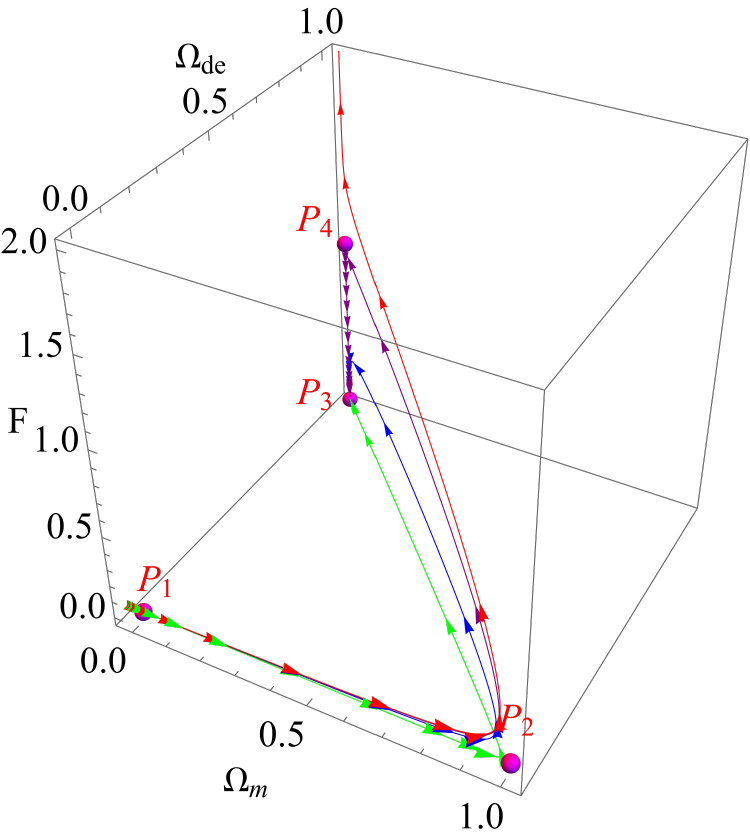}
\caption{\label{Fig7} Phase space trajectories in the $(\Omega_{m},\Omega_{de},F)$ space for the $\Lambda$CDM and FHDEF models, shown as green solid and red solid lines respectively, with mean parameter values taken from Table~\ref{Tab1}. The blue and purple solid lines correspond to the cases for the FHDEF model with $\mathcal{C}=0.3$ and $\mathcal{C}=0.8$, respectively.}
\end{center}
\end{figure}

To further investigate the evolution trajectories, we plot the phase space trajectories in the $(\Omega_{m},\Omega_{de},F)$ space for the FHDEF model in Fig.~\ref{Fig7}, with the mean parameter values from Table~\ref{Tab1}. In this figure, the green solid and red solid lines represent the $\Lambda$CDM and FHDEF models with the mean parameter values respectively, and the blue and purple solid lines correspond to the cases for the FHDEF model with $\mathcal{C}=0.3$ and $\mathcal{C}=0.8$, respectively. This figure shows that the value of $\mathcal{C}$ determines the behavior of the evolution trajectory on the $F$-axis. As $\mathcal{C}$ increases, the trajectory deviates further from the $(\Omega_{m},\Omega_{de})$ plane. If $\mathcal{C}$ takes the value less than $0.8$, the universe will eventually evolve into the stable attractor $P_{3}$, as shown by the blue and purple solid lines. According to the trajectory of the FHDEF model with the mean parameter values depicted by the red solid line, we can determine the evolution of the universe in the FHDEF model: the universe originates from the radiation dominated epoch $P_{1}$, then passes through the pressureless matter dominated epoch $P_{2}$ and the dark energy dominated epoch $P_{4}$ successively, and eventually evolves parallel to $F$-axis with $\Omega_{m}=0$ and $\Omega_{de}=1$. Therefore, in the FHDEF model, after passing through the $\Lambda$CDM-like evolutionary stages, the universe does not converge to $P_{3}$ but continues to evolve further, revealing a richer dynamical behavior beyond the standard $\Lambda$CDM cosmology.

\section{Conclusions} \label{sec:5}

Based on the fractional entropy derived from fractional quantum mechanics and taking the Hubble horizon as IR cutoff, FHDEH model has been proposed. When the model parameter $\alpha$ takes some special values, FHDEH can realize the late time acceleration of the universe. In this paper, we construct the FHDEF and FHDEP models by adopting the future event horizon and the particle horizon as the IR cutoff, respectively.

After using the SN, OHD, and DESI DR2 datasets to constrain the model parameters for the FHDEH, FHDEF, and FHDEP models, we obtain the mean values of these parameters, as well as $\chi^{2}_{min}$, $\Delta \chi^{2}_{min}$, $\Delta$AIC, and $\Delta$BIC for these models. Compared to the $\Lambda$CDM model, the minimum $\chi^{2}_{min}$ values for these FHDE models are all lower than that of the $\Lambda$CDM model, indicating that the FHDE models provide a marginally lower $\chi^{2}_{min}$ than $\Lambda$CDM for this data combination. Among the three FHDE models, FHDEH gives the smallest $\Delta$AIC and $\Delta$BIC values. The AIC values of FHDEH and FHDEF are slightly lower than that of the $\Lambda$CDM model, indicating a marginal preference according to the AIC criterion. The $\Delta BIC$ values for the FHDEH, FHDEF, and FHDEP models are $2.6$, $10.1$, and $14.2$, respectively, all higher than that of the $\Lambda$CDM model, reflecting a penalty for the extra parameters $\alpha$ and $\mathcal{C}$. These results indicate that the BIC strongly favors the $\Lambda$CDM model over FHDEF and FHDEP models, and only slightly disfavors the FHDEH model.

When the dataset SN+OHD+DESI DR2+CMB, which includes the CMB distance priors, is used to constrain the FHDE models, the FHDEH and FHDEP models are strongly disfavored by the data, as the values of $\Delta \chi^{2}_{min}$ exceed $200$. For the FHDEF model, the values of $\Delta \chi^{2}_{min}$, $\Delta AIC$, and $\Delta BIC$ are $-0.8$, $3.2$, and $14$, respectively, suggesting that the FHDEF model yields a slightly lower $\chi^{2}_{min}$ than $\Lambda$CDM model, but it is strongly disfavored by the BIC due to the penalty for extra parameters. However, our previous work suggests that introducing an interaction term to the FHDEH model can improve the fit to the data~\cite{Huang2026}. Therefore, extending the FHDEH and FHDEP models by including an interaction between pressureless matter and dark energy may provide a viable resolution to their incompatibility with CMB data.

Then, using the mean values obtained from the dataset SN+OHD+DESI DR2+CMB, we solve the corresponding dynamical equation and plot the evolutionary curves of cosmological parameters for the FHDE models. We find that the FHDEF model predicts evolutions of $\Omega_{de}$ and $\Omega_{m}$ that are nearly identical to those of $\Lambda$CDM across the entire cosmic history, but $q$ deviate from $\Lambda$CDM in the future, while the FHDEH and FHDEP models exhibit significant deviations.

To further investigate the evolution of the universe in the FHDEF model, we plot its phase space trajectories in the $(\Omega_{m},\Omega_{de})$ plane and the $(\Omega_{m},\Omega_{de},F)$ space using the mean parameter values from the dataset SN+OHD+DESI DR2+CMB. The evolution of the universe in the FHDEF model can be summarized as follows: the universe originates from the radiation dominated epoch $P_{1}$, then passes through the pressureless matter dominated epoch $P_{2}$ and the dark energy dominated epoch $P_{4}$ successively, and eventually evolves parallel to the $F$-axis with $\Omega_{m}=0$ and $\Omega_{de}=1$. The results show that, after passing through the $\Lambda$CDM-like evolutionary stages, the universe does not converge to $P_{3}$ but continues to evolve further, revealing a richer dynamical behavior beyond the standard $\Lambda$CDM cosmology.

\acknowledgments{This work was supported by the National Natural Science Foundation of China under Grant Nos. 12405081 and 11865018. }

\end{document}